\documentclass[12pt]{iopart}
\usepackage{graphicx}
\usepackage{amssymb}
\usepackage{epsfig,url}
\newcommand{\richConnectivity}{\ensuremath{\phi}}

\newcommand{\pearsonCoeff}{\ensuremath{\alpha}}
\newcommand{\probab}{\ensuremath{P}}

\newcommand{\adjacencyElement}{\ensuremath{a}}
\newcommand{\richClubSet}{\ensuremath{\mathcal R}}
\newcommand{\numberLinks}{\ensuremath{L}}

\begin{document}

\title[Structural constraints in complex networks]{Structural constraints in complex networks}
\author{S. Zhou$^1$ and R. J. Mondrag\'on$^2$}
\address{$^1$Department of Computer Science, University College
London\\
Ross Building, Adastral Park, Ipswich, IP5 3RE, United Kingdom\\
$^2$Department of Electronic Engineering, Queen Mary, University of
London\\Mile End Road, London, E1 4NS, United Kingdom}
\ead{s.zhou@adastral.ucl.ac.uk}

\begin{abstract}
We present a link rewiring mechanism to produce surrogates of a
network where both the degree distribution and the rich--club
connectivity are preserved. We consider three real networks, the
AS--Internet, the protein interaction and the scientific
collaboration. We show that for a given degree distribution, the
rich--club connectivity is sensitive to the degree--degree
correlation, and on the other hand the degree--degree correlation is
constrained by the rich--club connectivity. In particular, in the
case of the Internet, the assortative coefficient is always negative
and a minor change in its value can reverse the network's rich--club
structure completely; while fixing the degree distribution and the
rich--club connectivity restricts the assortative coefficient to such
a narrow range, that a reasonable model of the Internet can be
produced by considering mainly the degree distribution and the
rich--club connectivity. We also comment on the suitability of using
the maximal random network as a null model to assess the rich--club
connectivity in real networks.
\end{abstract}




\section{Introduction}

In graph theory the degree $k$ is defined as the number of links a
node has. The distribution of degree $\probab(k)$ provides a global
view of a network's structure and is one of the most studied
topological properties. Many complex networks are scale--free because
they exhibit a power-law degree distribution, i.e.~$\probab(k)\sim
k^{-\gamma}$, $\gamma>1$~\cite{Barabasi99, albert02, newman03a,
dorogovtsev03a, Pastor04, Maslov02a, wasserman94, boccalettia06}. A
more complete description of  a network's structure is obtained from
the joint degree distribution
$\probab(k,k')$~\cite{Dorogovtsev02,dorogovtsev03,Mahadevan06}, which
is the probability that a randomly selected link connects a node of
degree $k$ with a node of degree $k'$. The degree distribution can be
obtained from the joint degree distribution: $\probab(k) = (\bar{k}/
k)\sum_{k'} \probab(k, k')$, where $\bar{k}$ is the average degree.

The joint degree distribution characterises the degree--degree
correlation~\cite{newman02,newman03} between two nodes connected by a
link. But in practice $\probab(k,k')$ can be difficult to measure, in
particular for a finite--size and scale--free
network~\cite{Boguna04}. Nevertheless the degree--degree correlation
can be inferred from the average degree of the nearest neighbours of
$k$-degree nodes~\cite{dorogovtsev03, Pastor01, vazquez03}, which is
a projection of the joint degree distribution given by
\begin{equation}
k_{nn}(k)=  {\bar{k}\sum_{k'}k'P(k,k')\over kP(k)}.
\end{equation}
If the nearest-neighbours average degree $k_{nn}$ is an increasing
function of $k$ then the network is assortative, where nodes tend to
attach to alike nodes, i.e.~high--degree nodes to high--degree nodes
and low--degree nodes to low--degree nodes. If $k_{nn}(k)$ is a
decreasing function of $k$ then the network is disassortative, where
high--degree nodes tend to connect with low--degree nodes.
A network's degree--degree correlation, or mixing pattern, can also
be summarised by a single scalar called the assortativity coefficient
$\alpha$, $-1\le\pearsonCoeff\le1$~\cite{newman02},
\begin{equation}
\pearsonCoeff = {\numberLinks^{-1}\sum_{j>i} k_ik_j a_{ij} -
\left[\numberLinks^{-1} \sum_{j>i} \frac{1}{2}(k_i +
k_j)a_{ij}\right]^2 \over \numberLinks^{-1}\sum_{j>i}
\frac{1}{2}(k_i^2+k_j^2)a_{ij}-\left[\numberLinks^{-1}\sum_{j>i}\frac{1}{2}
(k_i+k_j)a_{ij}\right]^2 },
\end{equation}
where $L$ is the total number of links, $k_i$, $k_j$ are the degrees
of nodes $i$ and $j$, and $\adjacencyElement_{ij}$ is an element of
the network's adjacency matrix, where $\adjacencyElement_{ij}=1$ if
nodes $i$ and $j$ are connected by a link otherwise
$\adjacencyElement_{ij}=0$~\cite{Costa06}. For an uncorrelated
network $\pearsonCoeff=0$, for an assortative network
$\pearsonCoeff>0$  and for a disassortative network
$\pearsonCoeff<0$.

In some scale--free networks the best connected nodes, rich nodes,
tend to be very well connected between themselves. A rich--club is
the set of nodes $\richClubSet_{>k}$ with degrees larger than a given
degree $k$. The connectivity between members of the rich--club is
measure by the rich--club connectivity~\cite{Zhou04a}, which is
defined as the ratio of the number of links $E_{>k}$ shared by the
nodes in the set $\richClubSet_{>k}$ to the maximum possible number
of links that the rich nodes can share,
\begin{equation}
\richConnectivity(k) = {E_{>k}\over | \richClubSet_{>k}|
\cdot(|\richClubSet_{>k} | - 1)/2}={1\over | \richClubSet_{>k}|
\cdot(|\richClubSet_{>k} | - 1)}\sum_{i,j\in\richClubSet_{>k}}
\adjacencyElement_{ij}, \label{equation:richclub}
\end{equation}
where $|\richClubSet_{>k}|$ is the number of nodes in the set
$\richClubSet_{>k}$~\cite{Costa06, Colizza06}. The rich--club
connectivity as a function of the degree is a global property of a
network. The rich--club connectivity is a different projection of the
joint degree distribution~\cite{Colizza06},
\begin{equation}
\phi(k) = {
N\bar{k}\sum_{k'=k+1}^{k_{max}}\sum_{k''=k+1}^{k_{max}}P(k',k'')
\over [N\sum_{k'=k+1}^{k_{max}}P(k')]\cdot[N
\sum_{k'=k+1}^{k_{max}}P(k')-1]},
\end{equation}
where $N$ is the total number of nodes and $k_{max}$ is the maximum
degree in the network. The rich--club connectivity and the
degree--degree correlation are not trivially related.

Our motivation here is twofold. First to study if the description of
a network  using $\probab(k)$ and $\richConnectivity(k)$ produces a
reasonable model of a real network. We consider three real networks,
the AS-Internet, the protein interaction and the scientific
collaboration. Our approach is, from a real network, to create
surrogate networks with the same $\probab(k)$, or even the same
$\richConnectivity(k)$, as the original network, and then compare
properties of the surrogates with the original network.
Second, we are interested in the properties of the surrogates, in
particular the maximal random case of a network, as it has been used
as a `null model' to assess network properties.

\begin{figure}
\centerline{\psfig{figure=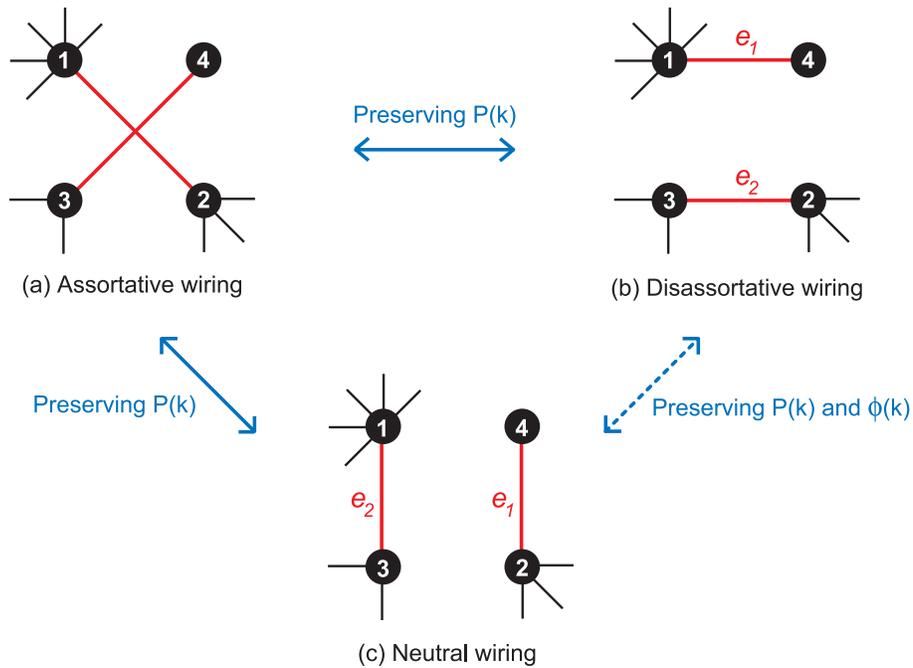,width=12cm}}
\caption{The four end nodes of a pair of links can be reconnected in
three wiring patterns: {\textbf{(a)}} assortative wiring, where one
link connects the two nodes with larger degrees and the other link
connects the two nodes with smaller degrees; {\textbf{(b)}}
disassortative wiring, where one link connects the node with the
highest degree with the node with the lowest degree and the other
link connects the two remaining nodes; and {\textbf{(c)}} neutral
wiring. \label{fig:rewiring}}
\end{figure}

\section{Link Rewiring Algorithms}

We create surrogate networks by using the link rewiring
algorithms~\cite{Maslov04,Xulvi04}.

\subsection{Maximal Cases I: Preserving $P(k)$}
The broad degree distribution $\probab(k)$ is an important
characteristic for complex networks and it should be preserved by any
link rewiring process~\cite{Newman01}. Figure~\ref{fig:rewiring}
shows that any four nodes with degrees $k_1\ge k_2\ge k_3\ge k_4$ can
be connected by two links in three possible wiring patterns. One can
see that reconnecting a pair of links from one wiring pattern to
another preserves the degree of individual nodes and therefore
preserves the degree distribution $P(k)$. It is possible to obtain
different kind of surrogate networks by rewiring links in the
following ways.

\begin{itemize}
\item \textbf{Maximal random case \emph{I}}: randomly choose a pair
of links and swap two of their end nodes. This is equivalent to
reconnect the four end nodes using a wiring pattern chosen at random.
The process is repeated for a sufficiently large number of times.

\item \textbf{Maximal assortative case \emph{I}}: reconnect a pair of
links in the assortative wiring pattern (see
figure~\ref{fig:rewiring}(a)) and repeat the process until all link
pairs are assortative wired.

\item \textbf{Maximal disassortative case \emph{I}}: similarly,
reconnect all pairs of links using the disassortative wiring pattern
(see figure~\ref{fig:rewiring}(b)).
\end{itemize}

For all link rewiring algorithms in this paper, a pair of links can
be rewired only if the resulted graph remains as a single connected
component.

\subsection{Maximal Cases II: Preserving Both $P(k)$ And $\phi(k)$}
It is possible to modify the link rewiring process such that the
rich--club connectivity  is preserved as well.
For a given degree $k$ the rich--club connectivity $\phi(k)$ depends
on the number of links $E_{>k}$ shared by the nodes belonging to the
set $\richClubSet_{>k}$. Any rewiring between nodes belonging to
$\richClubSet_{>k}$, or between nodes outside $\richClubSet_{>k}$,
will not change $E_{>k}$ hence $\richConnectivity(k)$ will remain the
same.
As shown in figure~\ref{fig:rewiring}, $E_{>k_1}$, $E_{>k_2}$,
$E_{>k_3}$ and $E_{>k_4}$ in the disassortative wiring
(figure~\ref{fig:rewiring}(b)) and the neutral wiring
(figure~\ref{fig:rewiring}(c)) are the same, because the link $e_1$
only and always belongs to  $E_{>k_4}$, and the other link $e_2$ only
and always belongs to $E_{>k_3}$ and $E_{>k_4}$.
This means that when reconnecting a pair of links between the
disassortative wiring and the neutral wiring, $\phi(k)$ remains
unchanged for all degrees. This allow us to obtain a different set of
maximal cases for a network while preserving both the network's
$P(k)$ and $\phi(k)$.
\begin{itemize}
\item \textbf{Maximal random case \emph{II}}: if a chosen pair of
links are assortatively wired, they are discarded and a new pair of
links is selected; otherwise the four end nodes are reconnected using
either the disassortative wiring or the neutral wiring at random.
\item \textbf{Maximal assortative case \emph{II}}: if a pair of links
are not assortatively wired, the four nodes are reconnected using the
neutral rewiring, which will produce a more assortative mixing than
using the disassortative wiring. The process is repeated for all
pairs of links.
\item \textbf{Maximal disassortative case \emph{II}}: if a pair of
links are not  assortatively wired, the four nodes are reconnected
using the disassortative wiring. The process is repeated for all
pairs of links.
\end{itemize}

\section{Results}

\begin{table}
\begin{center}
\caption{Three real networks considered are: (a)~the Internet network
at the autonomous system (AS)
level~\cite{Pastor04,Pastor01,faloutsos99,vazquez02,chen02,
mahadevan05b,zhou04d} from data collected by CAIDA~\cite{CAIDA}, in
which nodes represent Internet service providers and links
connections among those; (b)~the protein interaction
network~\cite{Maslov02a,colizza05} of the yeast \emph{Saccharomyces
cerevisiae} (http://dip.doe-mbi.ucla.edu/); and (c) the scientific
collaboration network~\cite{newman01a,newman01b}, in which nodes
represent scientists and a connection exists if they coauthored at
least one paper in the archive. The three networks contain multiple
components. In this paper we study the giant component of the
networks. We show the following properties: the number of nodes $N$
and links $L$ in the giant component, the average degree
$\bar{k}=2L/N$, the power--law exponent $\gamma$ by fitting $P(k)$
with $k^{-\gamma}$ for degrees between $6$ (the average degree) and
40, the maximum degree $k_{max}$, the power--law natural cut--off
degree $k_{cut}=N^{1/(\gamma-1)}$~\cite{Dorogovtsev02}, the
assortative coefficient $\pearsonCoeff$,  the rich--club
connectivity~$\phi(k_{>40})$, the rich--club exponent~$\theta$
obtained by fitting $\phi(k)$ with $k^{\theta}$ for degrees between
$6$ and 40, the size of rich--club clique $n_{clique}$, the average
shortest path length $\ell$, and the  average shortest path length
expected in a random graph $\ell^*=\ln
N/\ln{\bar{k}}$~\cite{Dorogovtsev02}. } \label{table:networks}
\renewcommand{\tabcolsep}{1.3pc} 
\renewcommand{\arraystretch}{1} 
\begin{tabular}{lrrr}
& & & \\
\hline\hline
& Internet &Protein      & Scientific     \\
&   & interaction     &  collaboration    \\
\hline
Number of nodes $N$&9,200& 4,626 & 12,722 \\
Number of links $L$&28,957& 14,801 & 39,967 \\
Average degree $\bar{k}$& 6.3& 6.4& 6.3\\
Power--law exponent $\gamma$ & 2.24 & 2.14 & 2.90 \\
Maximum degree $k_{max}$& 2,070& 282& 97\\
Power--law cut--off degree $k_{cut}$ & 1,573 & 1,641 & 145 \\
Assortative coefficient $\alpha$ & $-$0.236& $-$0.137 & 0.161 \\
Rich--club connectivity $\phi(k_{>40})$  & 26.8\% & 6.4\% & 7.1\%  \\
Rich--club exponent $\theta$ & 1.52  & 0.97 & 1.94 \\
Rich--club clique $n_{clique}$ & 16& 0 & 0 \\
Average shortest path length $\ell$ & 3.1 & 4.2 & 6.8 \\
Expected in a random graph $\ell^*$ & 5.0 & 4.5 & 5.1 \\
\hline\hline
\end{tabular}
\end{center}
\end{table}

\begin{figure}
\centerline{\psfig{figure=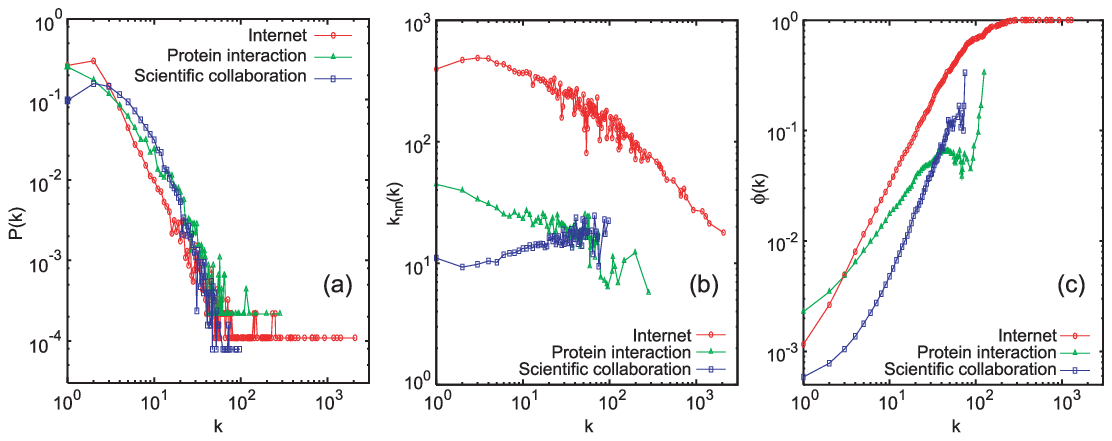,width=16cm}}
\caption{Topological properties of the Internet, the protein
interaction and the scientific collaboration networks: (a)~the degree
distribution, $P(k)$; (b)~the nearest-neighbours average degree of
$k$-degree nodes, $k_{nn}(k)$; and (c)~the rich--club connectivity as
a function of degree, $\phi(k)$. \label{fig:properties}}
\end{figure}

Table~1 describes the data sets and some of their topological
properties.
Figure~\ref{fig:properties}(a) shows that the three networks have a
power--law decay in $\probab(k)$. The degree distribution of the
Internet is well approximated  by $P(k)\sim k^{-\gamma}$,
$\gamma\simeq2.24$~\cite{Faloutsos99}, and it exhibits a fat tail
where the maximum degree, $k_{max}=2070$, is larger than the
power--law natural cut--off degree $k_{cut}=1573$. The degree
distribution of the protein interaction and the scientific
collaboration deviates from a strict power-law and have short tails.
Figure~\ref{fig:properties}(b) shows that the scientific
collaboration exhibits the assortative mixing behaviour, which is
common in social networks. The Internet and protein interaction
exhibit the disassortative mixing behaviour, which is typical for
technological and biological networks. The mixing behaviours are also
confirmed by evaluating the assortative coefficient of the networks
(see $\alpha$ in table~1).
Figure~\ref{fig:properties}(c) shows that the three data sets exhibit
different rich--club structures. Rich nodes in the disassortative
Internet are significantly more tightly interconnected with each
other than in the assortative scientific collaboration.  Only the
Internet contains a rich--club clique where the top 16 richest nodes
are fully connected with each other (see $n_{clique}$ in table~1).
One can see that an assortative network does not always exhibit a
strong rich-club structure, neither does a disassortative network
always lack a rich-club structure. Indeed high-degree nodes have very
large numbers of links and only a few of them are enough to provide
the connectivity to other high-degree nodes, whose number is anyway
small~\cite{Pastor04}.

A relevant metric of a network is the average shortest path length
$\ell$ between all nodes. As shown in table~1 the average shortest
path length in the Internet is significantly smaller than the average
shortest path length expected in a random graph with the same network
size. The Internet is so small~\cite{Cohen03} because it exhibits
both a strong rich-club structure and a strong disassortative mixing
behaviour. While members of the rich-club are tightly interconnected
with each other and they collectively function as a `super' traffic
hub for the network, the disassortative mixing ensures that the
majority of the network nodes, peripheral low-degree nodes, are
always near the rich-club core. Thus a typical shortest path between
two peripheral nodes consists of three hops, the first hop is from
the source node to a member of the rich-club, the second hop is
between two club members and the final hop is to the destination
node. One can see that a combination of the degree-degree correlation
and the rich-club connectivity can also explain the distribution of
the hierarchical path~\cite{Trusina04} and the short
cycles~\cite{Maslov04} in a network.

\begin{figure}
\centerline{\psfig{figure=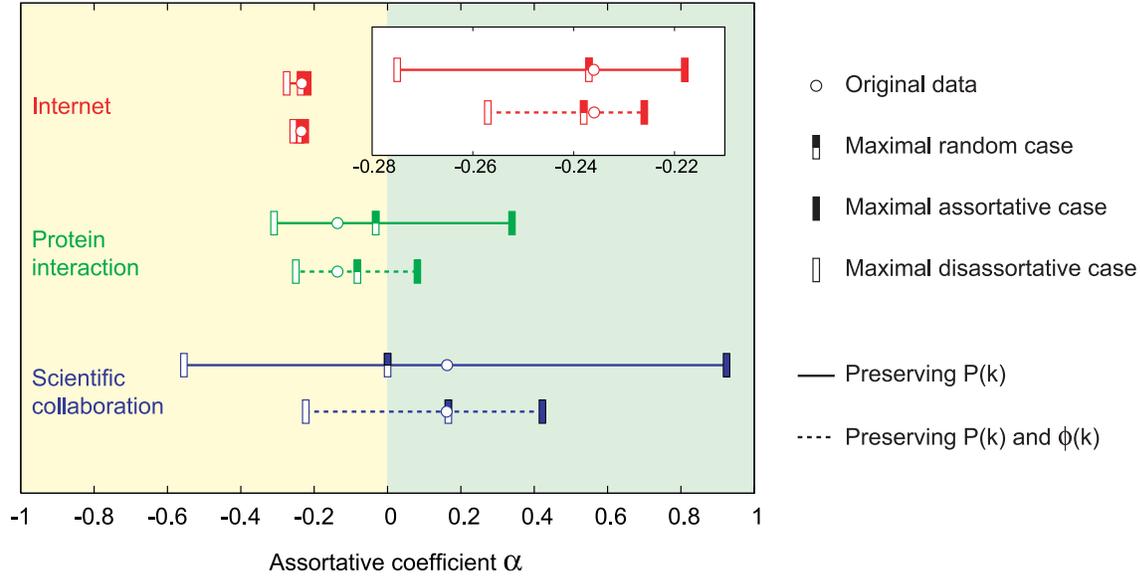,width=15cm}} \caption{
Range of the assortative coefficient $\alpha$ of the three networks
under study obtained by the link rewiring algorithms
preserving~$P(k)$ (case {\emph{I}}) comparing with that preserving
both $P(k)$ and $\phi(k)$ (case {\emph{II}}). The inset shows the
enlargement for the Internet. The standard deviation of $\alpha$ for
a maximal rewired case is smaller  than the symbol representing
it.\label{fig:alpha}}
\end{figure}

Figure~\ref{fig:alpha} shows the range of the assortative coefficient
$\pearsonCoeff$ obtained by the link rewiring algorithms preserving
the degree distribution (case~\emph{I}) against that preserving both
the degree distribution and the rich--club connectivity
(case~\emph{II}). The maximal random case of a real network is
averaged over 40 surrogate networks, each of which is obtained by
repeating the appropriate link rewiring process for $1000\times L$
times, where $L$ is the total number of links in the network.

For case~\emph{I} preserving $\probab(k)$, the maximal random
rewiring of the protein interaction and the scientific collaboration
almost decorrelates the networks, and the assortative and
disassortative rewiring can produce surrogate networks in the range
from assortative to disassortative. This is in contrast to the
Internet, where the maximal random case is almost as disassortative
as the original data. In fact all the surrogate networks produced by
rewiring the Internet are disassortative, the assortative coefficient
is always negative and its value is restricted to a very small range.
This behaviour of the Internet is due to the restriction of having a
finite network that has a power--law decay in its degree distribution
and that the maximum degree is larger than the natural cut--off
degree~\cite{Dorogovtsev02,Boguna04}.

For case~\emph{II} preserving both $P(k)$ and $\phi(k)$, the range of
$\pearsonCoeff$ is narrower than case~\emph{I} when only fixing
$P(k)$. This result confirms the analytical analysis by Krioukov and
Krapivsky~\cite{Krioukov05} that the rich--club connectivity
constrains a network's degree--degree correlation. In the case of the
Internet, the assortative coefficient is restricted to an even
smaller range. This observation suggests that a reasonable model of a
real network can be produced by modelling the degree distribution and
the rich--club connectivity, e.g.~the Positive-Feedback
Preference~(PFP) model~\cite{zhou04d,zhou06b,zhou06c} for the
Internet.

\begin{figure}
\centerline{\psfig{figure=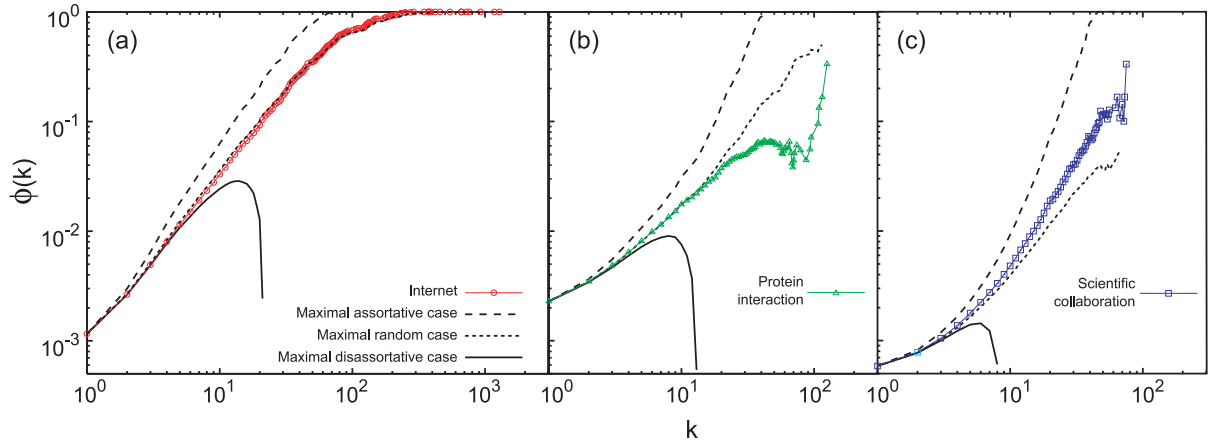,width=16cm}} \caption{
Rich--club connectivity $\phi(k)$ of (a)~the Internet, (b)~the
protein interaction, and (c) the scientific collaboration, comparing
with the three maximal cases \emph{I} obtained by preserving
$P(k)$.\label{fig:RichRewired}}
\end{figure}

Figure~\ref{fig:RichRewired} shows the rich--club connectivity of the
three networks each compared with their three maximal cases
\emph{{I}} obtained by preserving $P(k)$. The rich--club connectivity
changes dramatically due to the rewiring. For all the maximal
assortative networks there is a notable increase of~$\phi(k)$
throughout all degrees and all contain a fully connected rich--club
clique which consists nodes with degrees larger than 78, 48 and 46
for the Internet, the protein interaction and the scientific
collaboration respectively. For all the maximal disassortative
networks there is a complete collapse of the rich--club structure
such that there is no single link shared between nodes with degrees
larger than 23.
This suggests that networks with the same degree distribution can
have very different rich--club connectivity. In other words the
degree distribution does not constrain the rich--club connectivity.
The rich--club connectivity is  sensitive to the change of a
network's degree--degree correlation. For the Internet, a minor
change in the assortative coefficient within the narrow range of
$\pearsonCoeff\in(-0.218,-0.275)$ could reverse the rich--club
structure completely. This highlights the importance to measure the
rich--club connectivity when evaluating a network model.

\section{Discussion}

The maximal random network obtained by preserving $P(k)$ has been
used to discern whether  the existence of an interaction between two
proteins is due to chance or not~\cite{Maslov02a}. To do such, the
probability that two nodes share a link in the protein interaction
network is compared against the probability that the same two nodes
will share a link in the maximal random network. The maximal random
network is used as a null model because in this case it is almost a
decorrelated network (see figure~\ref{fig:alpha}).

Recently the maximal random network has also been used as a null
model to detect the origin of the rich--club connectivity in real
networks~\cite{Colizza06}. The argument is that if the rich--club
connectivity of the original network is the same as that of the
maximal random network then the rich--club connectivity was created
by chance, otherwise there was an `organisational principle'
responsible for the existence (or the lack) of the rich--club
structure. In the case of the Internet, the original
network\footnote{[19] used the Internet data collected by the Route
Views project [37], which exhibits a similar rich-club structure as
the Internet data collected by CAIDA used in this paper.} and the
maximal random network have similar rich-club connectivity (see
figure~\ref{fig:RichRewired}(a)), then the conclusion in
Ref.~\cite{Colizza06, Amaral06} was that `\emph{hubs in the Internet
... are not tightly interconnected}' and `\emph{the Internet does not
have an oligarchic structure whereas, for example, scientific
collaborations do}'. However, as shown in
figure~\ref{fig:properties}(c), the Internet does contain a well
connected rich-club core and we do not need more statistical analysis
to support this observation.

To understand the problem of using the maximal random network of the
Internet as a null model, one need to realise that the maximal random
network in this case is not an uncorrelated network. On the contrary
it is a strongly correlated network and is almost as dissasortative
as the Internet. Rich nodes in both the original network and the
maximal random network are tightly interconnected, and the similarity
between the rich--club connectivity of the two networks does not
implies that the Internet lacks a rich-club structure.

Notice that the maximal random network for the Internet with $P(k)$
fixed is more dissasortative than the original network, and the
maximal random network with $P(k)$ and $\phi(k)$ both fixed is even
more dissasortative (see inset in figure~3). This suggests that the
rich--club structure depends strongly on the nature of the
degree--degree correlation and it was not formed by chance. This
strong dependence on the tail of the degree distribution ($k_{max}$)
and the degree--degree correlation has also been noted in the
estimates of large cliques that appear in random scale--free
networks~\cite{Bianconi06}. A more detail analysis of the null--model
of the rich--club connectivity will be published elsewhere.

\section{Conclusions}
The rich--club connectivity and the degree--degree correlation
describe the global structure of a network from different
perspectives. We show that for a given degree distribution, the
rich--club connectivity is sensitive to the degree--degree
correlation, and on the other hand the degree--degree correlation is
constrained by the rich--club connectivity. In particular for the
case of the Internet, the assortative coefficient is always negative
and a minor change in its value can reverse the network's rich--club
structure completely; if fixing both the degree distribution and the
rich--club connectivity, the assortative coefficient is restricted to
such a narrow range that a reasonable model of the Internet can be
produced by considering mainly the degree distribution and the
rich--club connectivity.

We also clarify some misinterpretations that have appeared in the
literature which use the maximal random case as a null model to
assess the rich-club connectivity in real networks. We remark that
some care is needed to avoid reaching misleading conclusions, in
particular when studying the Internet.

\section*{Acknowledgments}

We thank A.~Vespignani and V.~Colizza for sharing the data sets of
the protein interaction and scientific collaboration networks. We
also thank CAIDA for providing the data set of the Internet AS graph.
SZ is partially funded by the UK Nuffield Foundation under grant
no.~NAL/01125/G, and RJM is partially funded by the UK EPSRC under
grant no.~EP/C520246/1.

\section*{References}


\begin{thebibliography}{10}

\bibitem{Barabasi99}
A.~Barab\'asi and R.~Albert.
\newblock Emergence of scaling in random networks.
\newblock {\em Science}, 286:509, 1999.

\bibitem{albert02}
R.~Albert and A.~L. Barab\'asi.
\newblock Statistical mechanics of complex networks.
\newblock {\em Rev. Mod. Phys.}, 74:47--97, 2002.

\bibitem{newman03a}
{M. E. J.} Newman.
\newblock The structure and function of complex networks.
\newblock {\em SIAM Review}, 45:167--256, 2003.

\bibitem{dorogovtsev03a}
S.~N. Dorogovtsev and J.~F.~F. Mendes.
\newblock {\em {Evolution of Networks - From Biological Nets to the Internet
  and WWW}}.
\newblock Oxford University Press, Oxford, 2003.

\bibitem{Pastor04}
R.~Pastor-Satorras and A.~Vespignani.
\newblock {\em Evolution and Structure of the Internet - A Statistical Physics
  Approach}.
\newblock Cambridge University Press, Cambridge, 2004.

\bibitem{Maslov02a}
S.~Maslov and K.~Sneppen.
\newblock Specificity and stability in topology of protein networks.
\newblock {\em Science}, 296(5569):910--913, 2002.

\bibitem{wasserman94}
S.~Wasserman and K.~Faust.
\newblock {\em Social Network Analysis}.
\newblock Cambridge University Press, Cambridge, 1994.

\bibitem{boccalettia06}
S.~Boccaletti, V.~Latora, Y.~Moreno, M.~Chavez, and D.-U. Hwang.
\newblock Complex networks: {S}tructure and dynamics.
\newblock {\em Physics Reports}, 424:175--308, 2006.

\bibitem{Dorogovtsev02}
S.~N. Dorogovtsev and J.~F.~F. Mendes.
\newblock Evolution of networks.
\newblock {\em Adv. Phys.}, 51(1079), 2002.

\bibitem{dorogovtsev03}
S.~N. Dorogovtsev.
\newblock Networks with given correlations.
\newblock \url{http://arxiv.org/abs/cond-mat/0308336v1}.

\bibitem{Mahadevan06}
P.~Mahadevan, D.~Krioukov, K.~Fall, and A.~Vahdat.
\newblock Systematic topology analysis and generation using degree
  correlations.
\newblock In {\em Proc.~of SIGCOMM'06}, pages 135--146. ACM Press, New York,
  2006.

\bibitem{newman02}
M.~E.~J Newman.
\newblock Assortative mixing in networks.
\newblock {\em Phys. Rev. Lett.}, 89(208701), 2002.

\bibitem{newman03}
M.~E.~J. Newman.
\newblock Mixing patterns in networks.
\newblock {\em Phys. Rev. E}, 67(026126), 2003.

\bibitem{Boguna04}
M.~{Bogu\~n\'a}, R.~Pastor-Satorras, and A.~Vespignani.
\newblock Cut-offs and finite size effects in scale-free networks.
\newblock {\em Eur. Phys. J. B}, 38:205--210, 2004.

\bibitem{Pastor01}
R.~Pastor-Satorras, A.~V\'azquez, and A.~Vespignani.
\newblock Dynamical and correlation properties of the internet.
\newblock {\em Phys. Rev. Lett.}, 87(258701), 2001.

\bibitem{vazquez03}
A.~V\'azquez, M.~{Bogu\~n\'a}, Y.~Moreno, R.~Pastor-Satorras, and
  A.~Vespignani.
\newblock Topology and correlations in structured scale-free networks.
\newblock {\em Phys. Rev. E}, 67(046111), 2003.

\bibitem{Costa06}
L.~da~F.~Costa, F.~A. Rodrigues, G.~Travieso, and P.~R.~Villas Boas.
\newblock Characterization of complex networks: A survey of measurements.
\newblock \url{http://arxiv.org/abs/cond-mat/0505185}.

\bibitem{Zhou04a}
S.~Zhou and R.~J. Mondrag\'on.
\newblock The rich-club phenomenon in the {I}nternet topology.
\newblock {\em IEEE Comm. Lett.}, 8(3):180--182, March 2004.

\bibitem{Colizza06}
V.~Colizza, A.~Flammini, M.~A. Serrano, and A.~Vespignani.
\newblock Detecting rich-club ordering in complex networks.
\newblock {\em Nature Physics}, 2:110--115, 2006.

\bibitem{Maslov04}
S.~Maslov, K.~Sneppenb, and A.~Zaliznyaka.
\newblock Detection of topological patterns in complex networks: correlation
  profile of the internet.
\newblock {\em Physica A}, 333:529--540, 2004.

\bibitem{Xulvi04}
R.~Xulvi-Brunet and I.~M. Sokolov.
\newblock Reshuffling scale-free networks: From random to assortative.
\newblock {\em Phys. Rev. E}, 70(066102), 2004.

\bibitem{Newman01}
M.~E.~J. Newman, S.~H. Strogatz, and D.~J. Watts.
\newblock Random graphs with arbitrary degree distributions and their
  applications.
\newblock {\em Phys. Rev. E}, 64(026118), 2001.

\bibitem{faloutsos99}
M.~Faloutsos, P.~Faloutsos, and C.~Faloutsos.
\newblock On power--law relationships of the {I}nternet topology.
\newblock {\em Comput. Commun. Rev.}, 29:251--262, 1999.

\bibitem{vazquez02}
A.~V\'azquez, R.~Pastor-Satorras, and A.~Vespignani.
\newblock {Large-scale topological and dynamical properties of Internet}.
\newblock {\em Phys. Rev. E}, 65(066130), 2002.

\bibitem{chen02}
Q.~Chen, H.~Chang, R.~Govindan, S.~Jamin, S.~J. Shenker, and
W.~Willinger.
\newblock {The Origin of Power Laws in Internet Topologies (Revisited)}.
\newblock In {\em Proc.~of {INFOCOM} 2002}, pages 608--617. IEEE Computer
  Society, Washington D.C., 2002.

\bibitem{mahadevan05b}
P.~Mahadevan, D.~Krioukov, M.~Fomenkov, B.~Huffaker,
X.~Dimitropoulos,
  K.~Claffy, and A.~Vahdat.
\newblock {The Internet AS-level Topology: Three Data Sources and One
  Definitive Metric}.
\newblock {\em Comput. Commun. Rev.}, 36(1):17--26, 2006.

\bibitem{zhou04d}
S.~Zhou and R.~J. Mondrag\'on.
\newblock Accurately modelling the {I}nternet topology.
\newblock {\em Phys. Rev. E}, 70(066108), December 2004.

\bibitem{CAIDA}
{The Cooperative Association For Internet Data Analysis}.
\newblock \url{http://www.caida.org/}.

\bibitem{colizza05}
V.~Colizza, A.~Flammini, A.~Maritan, and A.~Vespignani.
\newblock Characterization and modeling of protein--protein interaction
  networks.
\newblock {\em Physica A}, 352:1--27, 2005.

\bibitem{newman01a}
{M. E. J.} Newman.
\newblock {Scientific collaboration networks. I. Network construction and
  fundamental results}.
\newblock {\em Phys. Rev. E}, 64(016131), 2001.

\bibitem{newman01b}
{M. E. J.} Newman.
\newblock {Scientific collaboration networks. II. Shortest paths, weighted
  networks, and centrality}.
\newblock {\em Phys. Rev. E}, 64(016132), 2001.

\bibitem{Cohen03}
R.~Cohen and S.~Havlin.
\newblock Scale-free networks are ultrasmall.
\newblock {\em Phys. Rev. Lett.}, 90(058701), 2003.

\bibitem{Trusina04}
A.~Trusina, S.~Maslov, P.~Minnhagen, and K.~Sneppen.
\newblock Hierarchy measures in complex networks.
\newblock {\em Phys. Rev. Lett.}, 92(17), 2004.

\bibitem{Krioukov05}
D.~Krioukov and P.~Krapivsky.
\newblock {Power Laws as a pre-asymptotic regime of the PFP Model}.
\newblock
  \url{http://www.caida.org/publications/presentations/2006/isma0605_dima/},
  2006.

\bibitem{zhou06b}
S.~Zhou.
\newblock {Understanding the evolution dynamics of Internet topology}.
\newblock {\em Phys. Rev. E}, 74(016124), 2006.

\bibitem{zhou06c}
S.~Zhou, G.-Qiang Zhang, and G.-Qing Zhang.
\newblock {The Chinese Internet AS-Level Topology}.
\newblock \url{http://arxiv.org/abs/cs.NI/0511101}.

\bibitem{oregon}
{Route Views Project}, {University of Oregon, Eugene}.
\newblock \url{http://www.routeviews.org/}.

\bibitem{Amaral06}
L.~A.~N. Amaral and R.~Guimer{\`a}.
\newblock Lies, damned lies and statistics.
\newblock {\em Nature Physics}, 2:75--76, 2006.

\bibitem{Bianconi06}
G.~Bianconi and M.~Marsili.
\newblock Emergence of large cliques in random scale-free networks.
\newblock {\em Europhys. Lett.}, 74(4):740, 2006.

\end{thebibliography}

\end{document}